\documentclass[twocolumn,showpacs,amsmath,amssymb,prl]{revtex4}

\usepackage{graphicx}
\usepackage{dcolumn}
\usepackage{bm}

\begin{document}
\title{
Spin current and accumulation generated by spin Hall insulator
}
\author{Masaru Onoda$^{1,3}$}
\email{m.onoda@aist.go.jp}
\author{Naoto Nagaosa$^{1,2,3}$}
\email{nagaosa@appi.t.u-tokyo.ac.jp}
\affiliation{
$^1$Correlated Electron Research Center (CERC),
National Institute of Advanced Industrial Science and Technology (AIST),
Tsukuba Central 4, Tsukuba 305-8562, Japan\\
$^2$Department of Applied Physics, University of Tokyo,
Bunkyo-ku, Tokyo 113-8656, Japan\\
$^3$CREST, Japan Science and Technology Corporation (JST),
Saitama, 332-0012, Japan
}
\begin{abstract}
Spin current and accumulation generated by the electric field 
in spin Hall insulator (SHI) are investigated theoretically 
in terms of the Keldysh formalism. In contrast to the quantum Hall system, 
there are no massless edge modes in general. 
The spin current is generated near the contacts to the electrodes by 
the hybridization between the metallic states and the conduction/valence 
bands of the SHI, but is truncated by the sink and source of the spin.
However, one can produce the spin current flowing out to the conductors,
which is attached to the SHI, and also the spin accumulation there
due to the leakage charge current
which breaks the time-reversal symmetry.
\end{abstract}
\maketitle
Intrinsic spin Hall effect (SHE) is an issue of recent intensive 
interests since it might offer an efficient way to inject the spins 
into the semiconductors without using magnets/magnetic field at room 
temperature with low energy cost~\cite{MNZ,MNZ1,Sinova}. 
Especially in the hole doped case~\cite{MNZ,MNZ1},
this intrinsic SHE is generated by the Berry curvature of 
the Bloch wavefunction in the momentum space.
Hence the spin current 
induced by this topological property is dissipationless in nature, 
similar to the charge current in quantum Hall effect (QHE).
However the voltage drop is produced by the ohmic charge current,
which produces the Joule heating.  Therefore the crucial question to 
address is `` Is there any system showing the spin Hall current 
without any dissipation analogous to the quantum Hall system (QHS)?''
Related to this question, it has been proposed that 
even band insulators could produce the nonzero SHE, 
i.e., spin Hall insulator (SHI)~\cite{SHI}. 
The idea there is that when the spin-dependent Berry curvature 
has the opposite sign for the conduction and valence bands, 
the occupied valence bands leads to the finite spin Hall effect. 
This is realized when the band gap is opened by 
the spin-orbit interaction (SOI) as in 
the case of HgTe, $\alpha$-Sn, and PbTe~\cite{SHI}. 
Also the quantum spin Hall effect (QSHE) is proposed for graphene~\cite{Kane}, 
where the up and down spins are decoupled into two quantum Hall systems.
In the generic cases, however, the (pseudo) spin is not conserved 
in the presence of SOI~\cite{Culcer,PZhang}
and the definition of the spin current is also obscure.
In refs.~\cite{MNZ1,SHI}, the conserved spin current is defined 
by projecting the spin operator into each of the band,
but this definition can not be applied for the spatially dependent 
situation in the presence of the disorder and/or the edges.
In refs.~\cite{Culcer,PZhang}, the another definition of the spin current
is given which includes the torque dipole density,
but the procedure of fixing the gauge ambiguity
is not clear in the present situation.
On the other hand, the spin and spin current is well-defined
in conductors without SOI attached to the SHI. 
Therefore it is worthwhile to pursue the spin current and
accumulation in the SHI by taking 
into account the realistic experimental setups with the 
edges and the contact to the electrodes/conductors. 
Especially it is important to clarify the difference between the SHI 
and the QHS including the role of edge modes. 
Here we summarize the several important differences between SHI and QHS: 
(i) there is no Chern-number characterizing the SHE in contrast to the QHE, 
(ii) the spin current is not conserved due to the SOI
while the charge current is always conserved, 
(iii) the effective theory of QHE contains
the U(1) Chern-Simons term, which dictates the existence of the 
massless edge modes to recover the local gauge invariance for finite 
size sample~\cite{Wen}, while there is no such a machinery in SHE of SHI. 
We will see below how these differences affect the low energy properties 
of SHI.

\begin{figure}[hbt]
\includegraphics[scale=0.25]{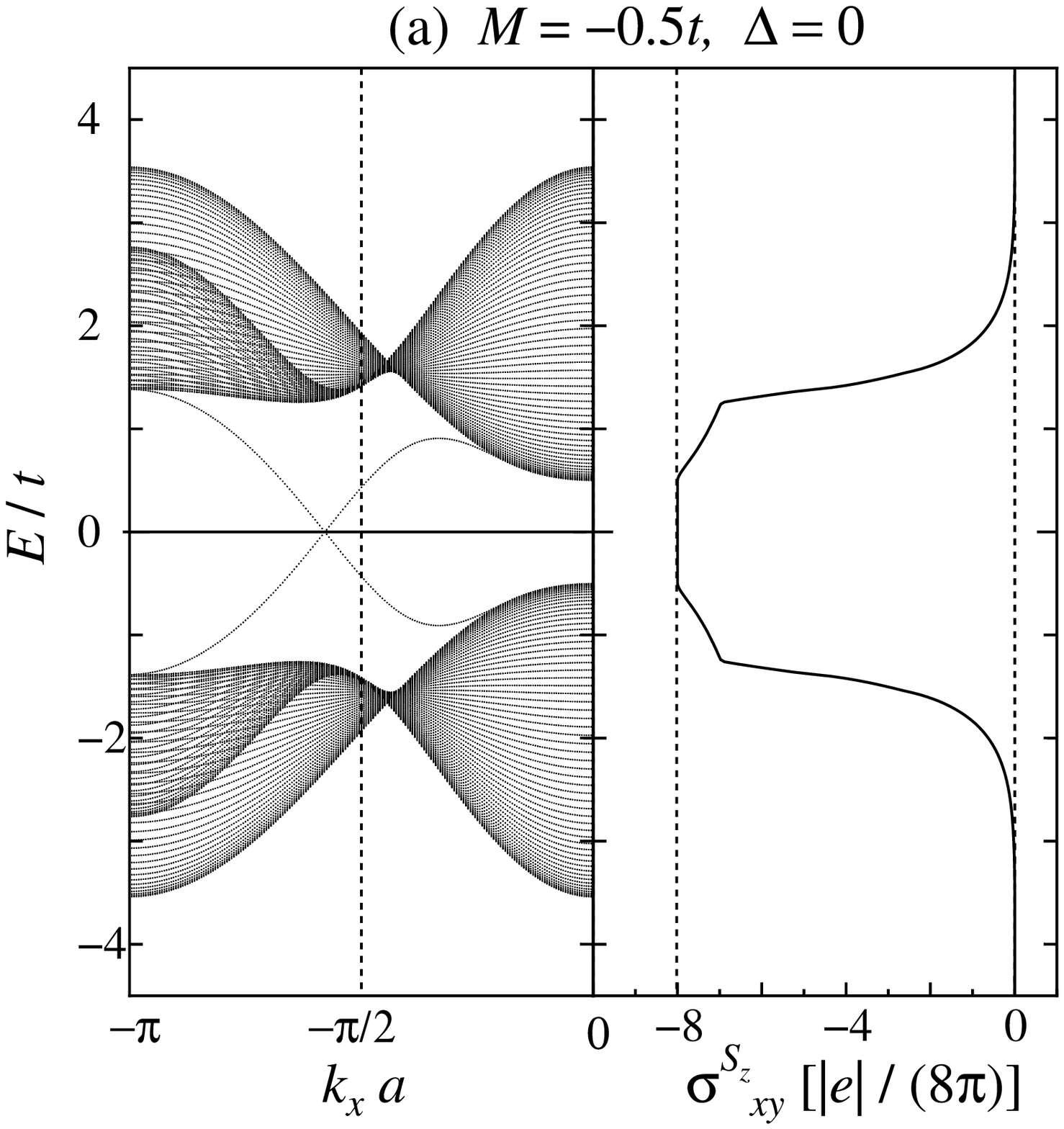}
\includegraphics[scale=0.25]{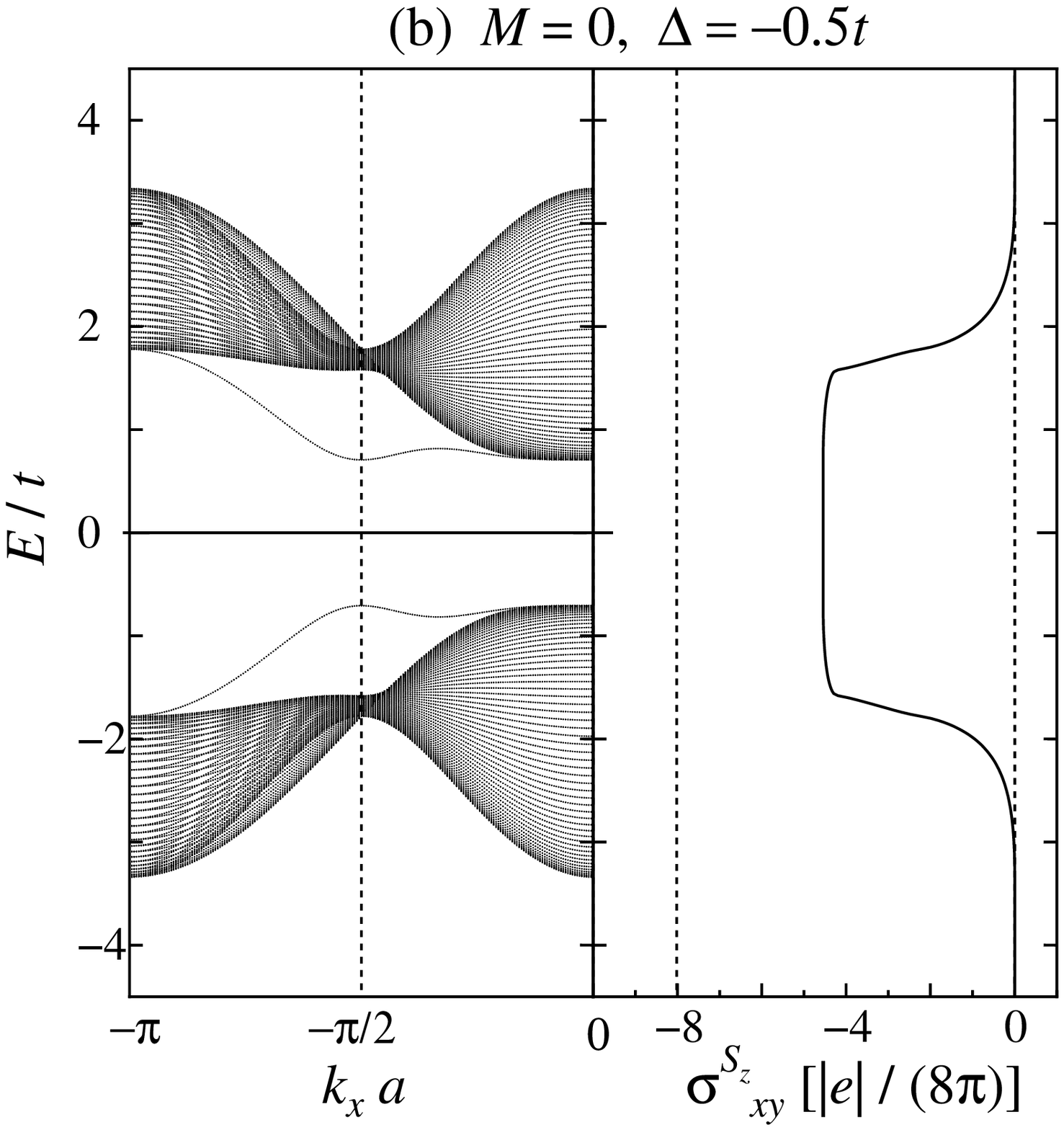}
\caption{
Energy dispersion of the semi-infinite system
and the bulk spin Hall conductance of the periodic system
for (a) the QSHS, and (b) the SHI.
Each curves in the energy dispersion corresponds to 
different $k_y$-value and the dispersion is symmetric between 
$k_x$ and $-k_x$.
} 
\label{fig:dis-ssig}
\end{figure}

We start with the following 4-band model for two-dimensional SHI 
defined on the lattice
and take the units in which $\hbar = c = 1$.
\begin{eqnarray}
H 
&=& \sum_{\bm{r},\bm{r}'}\sum_{\mu=1}^{5} 
c^{\dagger}_{\bm{r}}t^{\mu}_{\bm{r}\bm{r}'}\Gamma_{\mu}c_{\bm{r}'}
,\\
t^{1,2}_{\bm{r}\bm{r}'} 
&=& 
\begin{array}{rl}
\Delta, & \bm{r} = \bm{r}'
\end{array}
,\\
t^{3}_{\bm{r}\bm{r}'} 
&=& 
\left\{
\begin{array}{rl}
-\frac{t}{2\sqrt{2}}, & \bm{r} = \bm{r}'\pm (a \bm{e}_{x} + a \bm{e}_{y})\\
\frac{t}{2\sqrt{2}}, & \bm{r} = \bm{r}'\pm (a \bm{e}_{x} - a \bm{e}_{y})
\end{array}
\right.
,\\
t^{4}_{\bm{r}\bm{r}'} 
&=& \left\{
\begin{array}{rl}
-\frac{t}{\sqrt{2}}, & \bm{r} = \bm{r}' \pm a \bm{e}_{x}\\
\frac{t}{\sqrt{2}}, & \bm{r} = \bm{r}' \pm a \bm{e}_{y}
\end{array}
\right.
,\\
t^{5}_{\bm{r}\bm{r}'} 
&=& \left\{
\begin{array}{rl}
-\frac{t}{\sqrt{6}}, & \bm{r} = \bm{r}' \pm a \bm{e}_{x,y}\\
\frac{t}{\sqrt{6}}, & \bm{r} = \bm{r}' \pm (a \bm{e}_{x} \pm a \bm{e}_{y})\\
M, & \bm{r} = \bm{r}'
\end{array}
\right.
,\\
\end{eqnarray}
where $\Gamma$-matrices are defined in ref.~\cite{MNZ1}. 
Let us first consider the case of $\Delta= 0$,
where the $4\times 4$ $\Gamma$-matrix space
will be decoupled to two of the $2\times 2$ matrix space.
Correspondingly, the bands are split into  the electron- and hole-like bands
and each of them is still doubly degenerate.
When $\mathrm{sgn}(M)=\mathrm{sgn}(t)$,
all bands have zero Chern-number,
although they have non-zero Berry curvature distribution in momentum space.
Therefore, edge modes do not disperse across the bulk gap.
(It is noted that the gap in the edge mode can be narrower than the bulk gap.)
In the case of $\mathrm{sgn}(M)=-\mathrm{sgn}(t)$,
each band has non-zero Chern-number ($\pm 2$)
and there appear edge modes across the bulk gap ($=2|M|$) 
just like the QHS as shown in Fig.~\ref{fig:dis-ssig}~(a).
It is interesting that each state near the $\Gamma$-point
in Brillouin zone is described as 
a generalized Dirac-fermion with quadratic dispersion~\cite{Onoda}.
The spin Hall conductivity is quantized in the clean limit,
when the Fermi energy is in the bulk gap.
This corresponds to the decoupled two QHS's with
opposite sign of the Hall conductance
and we shall call this case as the quantum spin Hall system (QSHS).
However, it is noted that there is a crucial difference 
between our QSHS and that in ref.~\cite{Kane}.
In our QSHS, any spin components are not conserved,
although the spin $z$-component is conserved in the QSHS of ref.~\cite{Kane}.
Therefore, we assume the spin-independent chemical potential
even in a non-equilibrium state.
This treatment means the scattering
between edge modes with opposite chiralities on a same edge.
Therefore, in contrast to the QHS and the QSHS in ref.~\cite{Kane},
there appears the voltage drop in the edge modes of our QSHS.

Now we consider the case $\Delta \neq 0$, which causes 
the hybridization between previously separated two $2\times2$ matrices. 
This is usually the case in the three dimensional model because
kinetic terms contains $\Gamma_{1,2}$-matrices.
Correspondingly the hybridization of the edge modes
with opposite chiralities occurs causing the real gap opening 
as shown in Fig.~\ref{fig:dis-ssig}~(b).
When $M=0$, the bulk gap is estimated as $2\sqrt{2}|\Delta|$.

Here, we employ the Keldysh formalism applied to the finite size sample 
attached to the electrodes~\cite{Datta}.
Recently this formalism is applied to the 2-band system with Rashba coupling~\cite{Nikolic}.
The real-space Green's 
functions are used to obtain the charge/spin current and spin accumulation.
The effect of electrodes is incorporated by the contact self-energy.
In this formalism, the lesser Green function $G^{<}$ is the most 
essential correlation function to compute physical observables, 
e.g. charge and spin densities, charge and spin current densities. 
$G^{<}$ is defined by
$
G^{<}_{[\bm{r}\sigma][\bm{r}'\sigma']}(t,t') 
= i\langle c^{\dagger}_{\bm{r}'\sigma'}(t')c_{\bm{r}\sigma}(t)\rangle
$,
where $c^{(\dagger)}_{\bm{r}\sigma}(t)$ is the annihilation(creation) operator of 
the electron with quantum index $\sigma$ at the position $\bm{r}$ and the time $t$.
For simplicity, we shall focus on the steady state, i.e.,
$G^{<}_{[\bm{r}\sigma][\bm{r}'\sigma']}(t,t') = G^{<}_{[\bm{r}\sigma][\bm{r}'\sigma']}(t-t')$,
and take the Fourier transformation with respect to the time index, i.e.,
$G^{<}_{[\bm{r}\sigma][\bm{r}'\sigma']}(t-t')\to G^{<}_{[\bm{r}\sigma][\bm{r}'\sigma']}(E)$.
The expectation value of observable
$\hat{\mathcal{O}}_{\bm{r}'\bm{r}} = c^{\dagger}_{\bm{r}'}\mathcal{O}_{\bm{r}'\bm{r}} c_{\bm{r}}$ is given by
\begin{equation}
\langle\hat{\mathcal{O}}_{\bm{r}'\bm{r}}\rangle = 
-i\int \frac{dE}{2\pi}\:\mathrm{Tr}^{(\sigma)}[\mathcal{O}_{\bm{r}'\bm{r}}G^{<}_{\bm{r}\bm{r}'}(E)],
\end{equation}
where $\mathrm{Tr}^{(\sigma)}$ means the trace with respect to the index $\sigma$.

In order to obtain $G^{<}(E)$, we solve the following equations, 
$G^{<}(E) = G^{R}(E)\Sigma^{<}(E)G^{A}(E)$,
$[E-H-\Sigma^{R(A)}(E)]G^{R(A)} = I$,
where 
$\Sigma^{<}(E)$ the lesser self-energy,
$G^{R(A)}(E)$ the retarded(advanced) Green function,
and $\Sigma^{R(A)}(E)$ the retarded(advanced) self-energy.
We employ the following approximation for the self-energy 
to incorporate the non-equilibrium 
nature of the system and the effects of electrodes and static disorders.
\begin{equation}
\Sigma^{R}_{[\bm{r}\sigma][\bm{r}'\sigma]}(E) 
= \Sigma^{R,\mathrm{cont}}_{[\bm{r}\sigma][\bm{r}'\sigma]}(E)
+\frac{-i}{2\tau_{\bm{r}}(E)}\delta_{[\bm{r}\sigma][\bm{r}'\sigma']},
\end{equation}
where
$\Sigma^{R,\mathrm{cont}}(E)$ is the contact self-energy,
and $\tau_{\bm{r}}(E)$ the local lifetime due to disorders.
The local lifetime $\tau_{\bm{r}}(E)$ is determined self-consistently 
by the recursion equation,
$
\frac{1}{\tau_{\bm{r}}(E)} = \gamma N_{\bm{r}}(E)
$,
where $\gamma$ represents the strength of disorder
and $N_{\bm{r}}(E)=\frac{i}{(2\pi)}
\mathrm{Tr}^{(\sigma)}[G^{R}_{\bm{r}\bm{r}}(E)-G^{A}_{\bm{r}\bm{r}}(E)]$ 
is the local density of states per unit cell.
This corresponds to the self-consistent Born approximation in real space. 
The non-equilibrium nature is incorporated through the local 
distribution function $f_{\bm{r}}(E)$ as
\begin{equation}
\Im\left[\Sigma^{<}_{[\bm{r}\sigma][\bm{r}'\sigma]}(E)\right]
= \left[
\frac{1}{\tau^{\mathrm{cont}}_{\bm{r}}(E)} 
+\frac{1}{\tau_{\bm{r}}(E)}\right]
f_{\bm{r}}(E)\delta_{[\bm{r}\sigma][\bm{r}'\sigma']},
\end{equation}
where $\frac{1}{\tau^{\mathrm{cont}}(E)}=-2\Im[\Sigma^{R,\mathrm{cont}}(E)]$.
$f_{\bm{r}}(E)$ is determined self-consistently by the recursion equation
$
n_{\bm{r}}(E)=\frac{1}{2\pi}\mathrm{Tr}^{(\sigma)}[G^{<}_{\bm{r}\bm{r}}(E)] 
= N_{\bm{r}}(E)f_{\bm{r}}(E)
$,
with the boundary condition at the contacts
$
f_{\bm{r}}(E)|_{n \mathrm{th\: contact}} = \frac{1}{1+e^{\beta(E-\mu_{n})}}
$,
where $\mu_{n}$ is the chemical potential at the $n$-th contact.
It should be noted that the above approximation automatically satisfies
the charge conservation except for the contacts with electrodes where
there may be incoming or outgoing charge current.
The higher-level correction over the self-consistent Born approximation
is taken into account through the recursion equation for $f_{\bm{r}}(E)$
which contains the vertex correction of the external electric field.
Therefore the Ward-Takahashi identity holds in the above approximation.

\begin{figure}[hbt]
\includegraphics[scale=0.18]{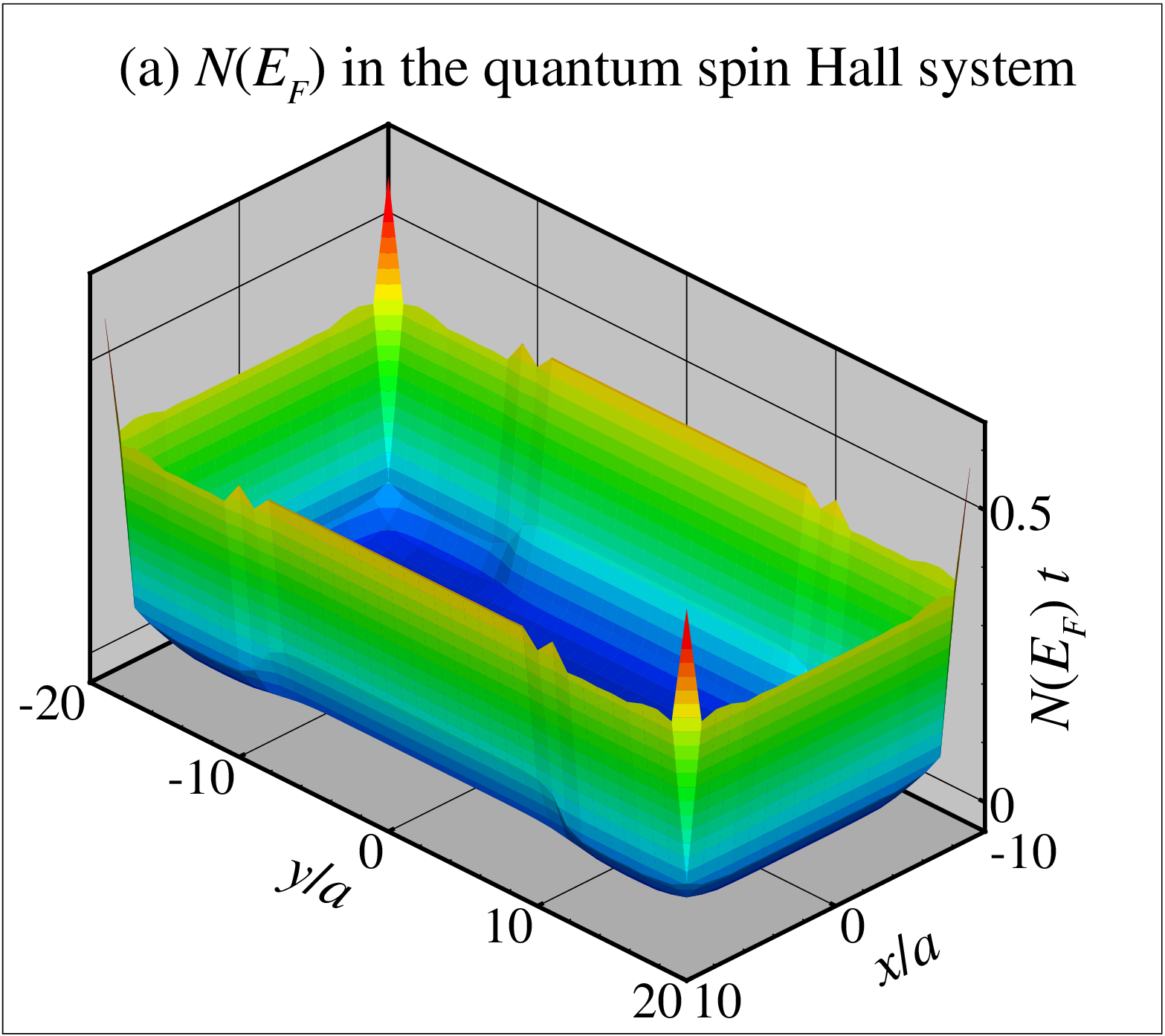}
\includegraphics[scale=0.18]{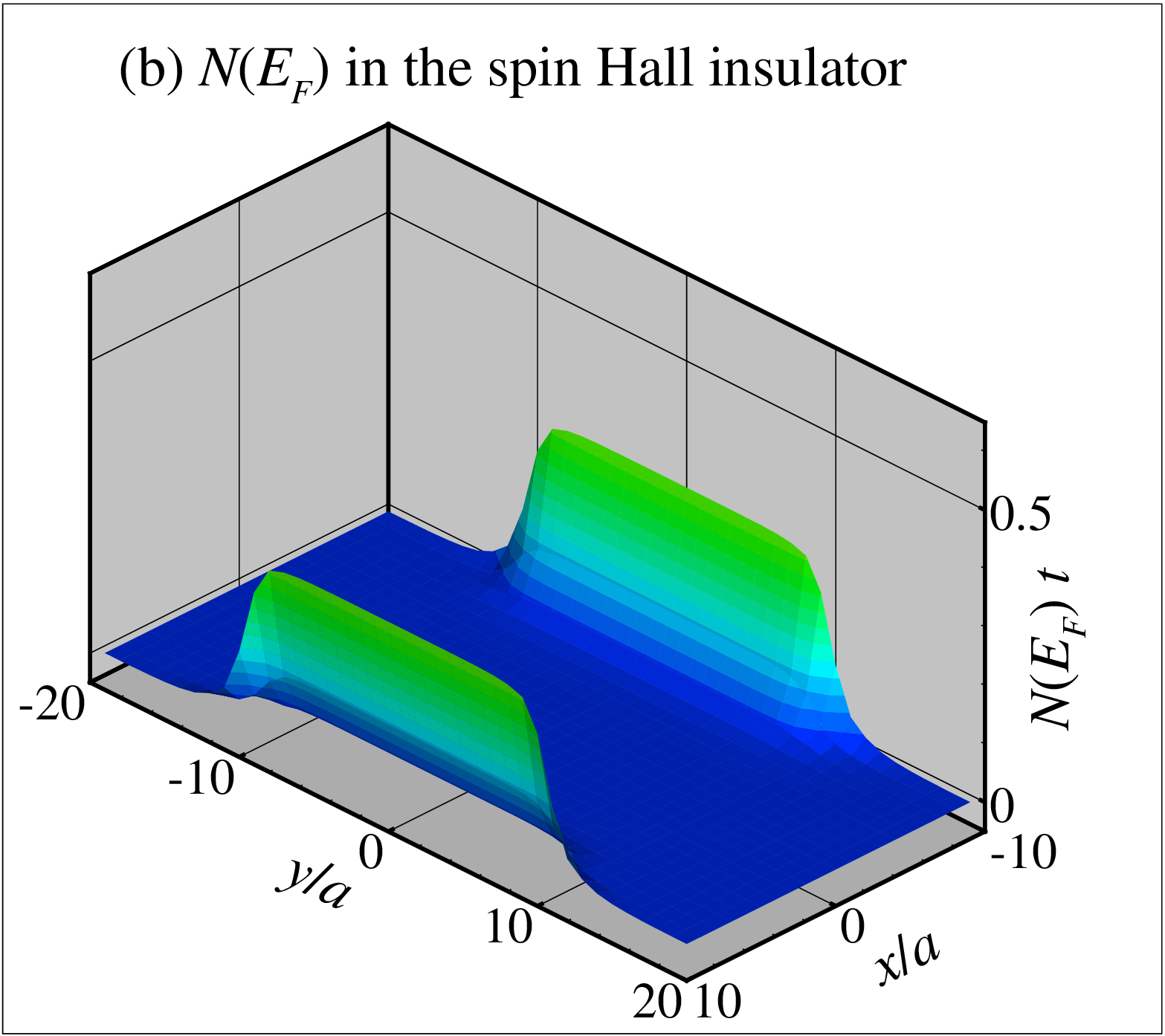}
\includegraphics[scale=0.18]{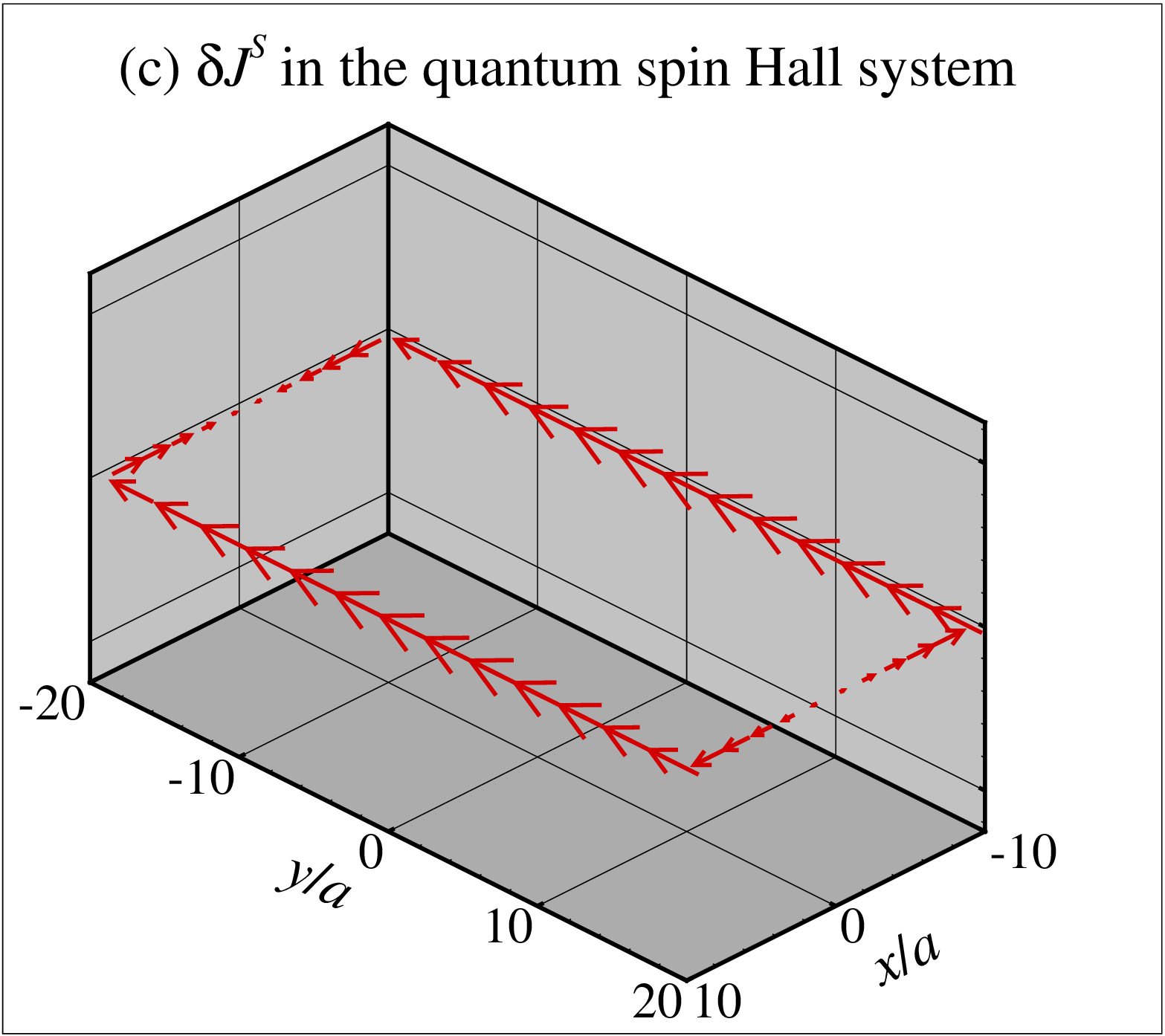}
\includegraphics[scale=0.18]{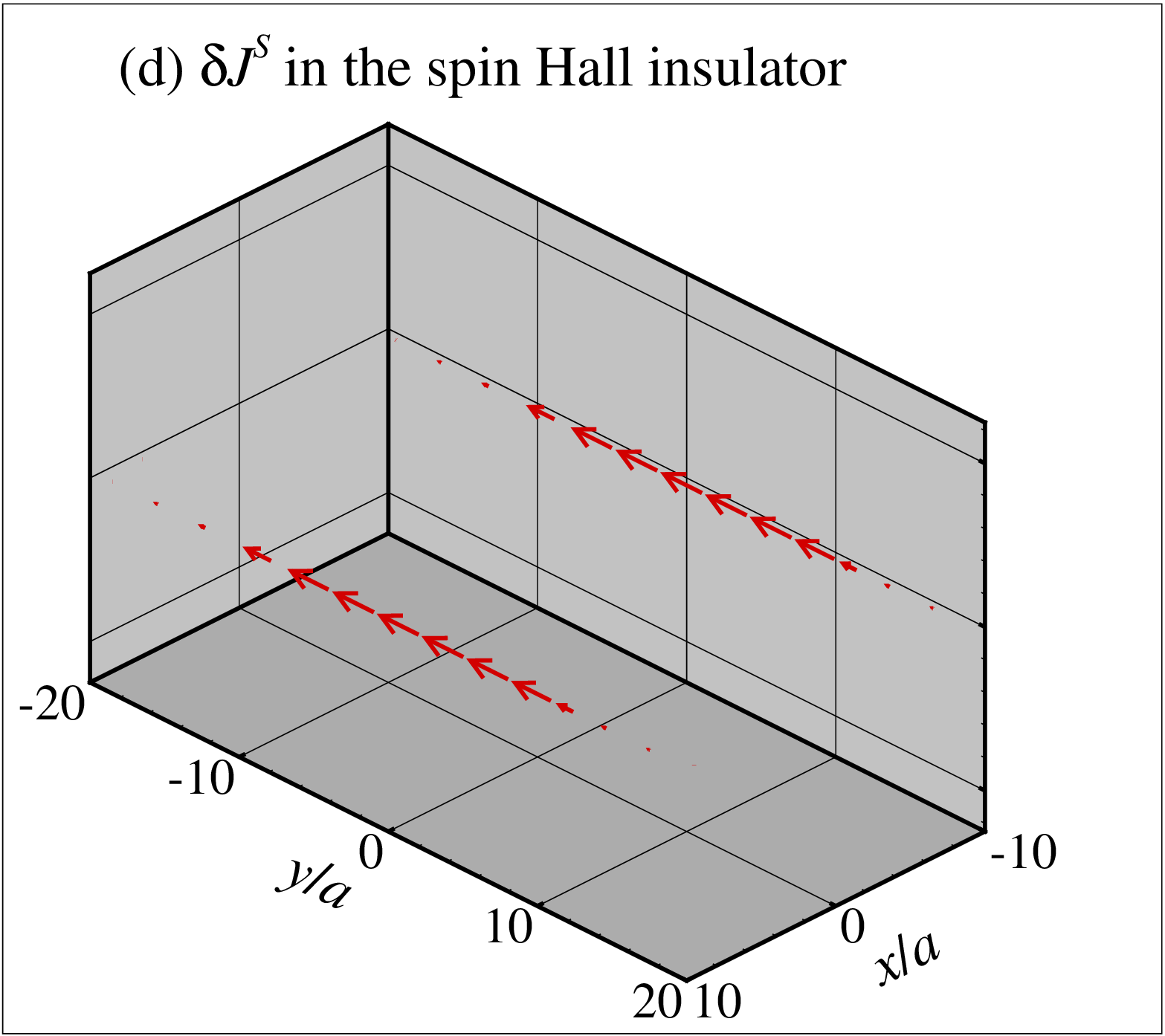}
\includegraphics[scale=0.18]{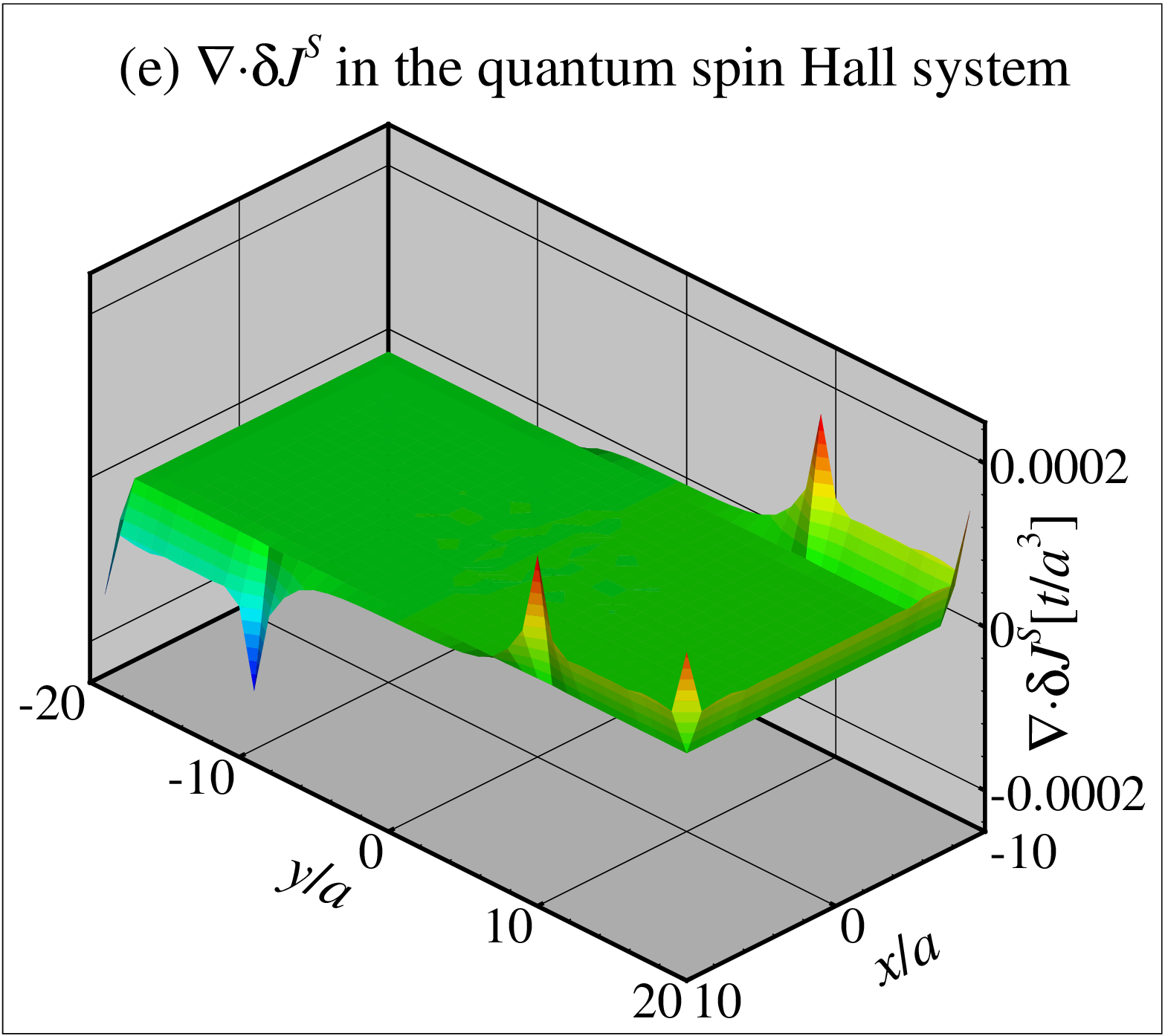}
\includegraphics[scale=0.18]{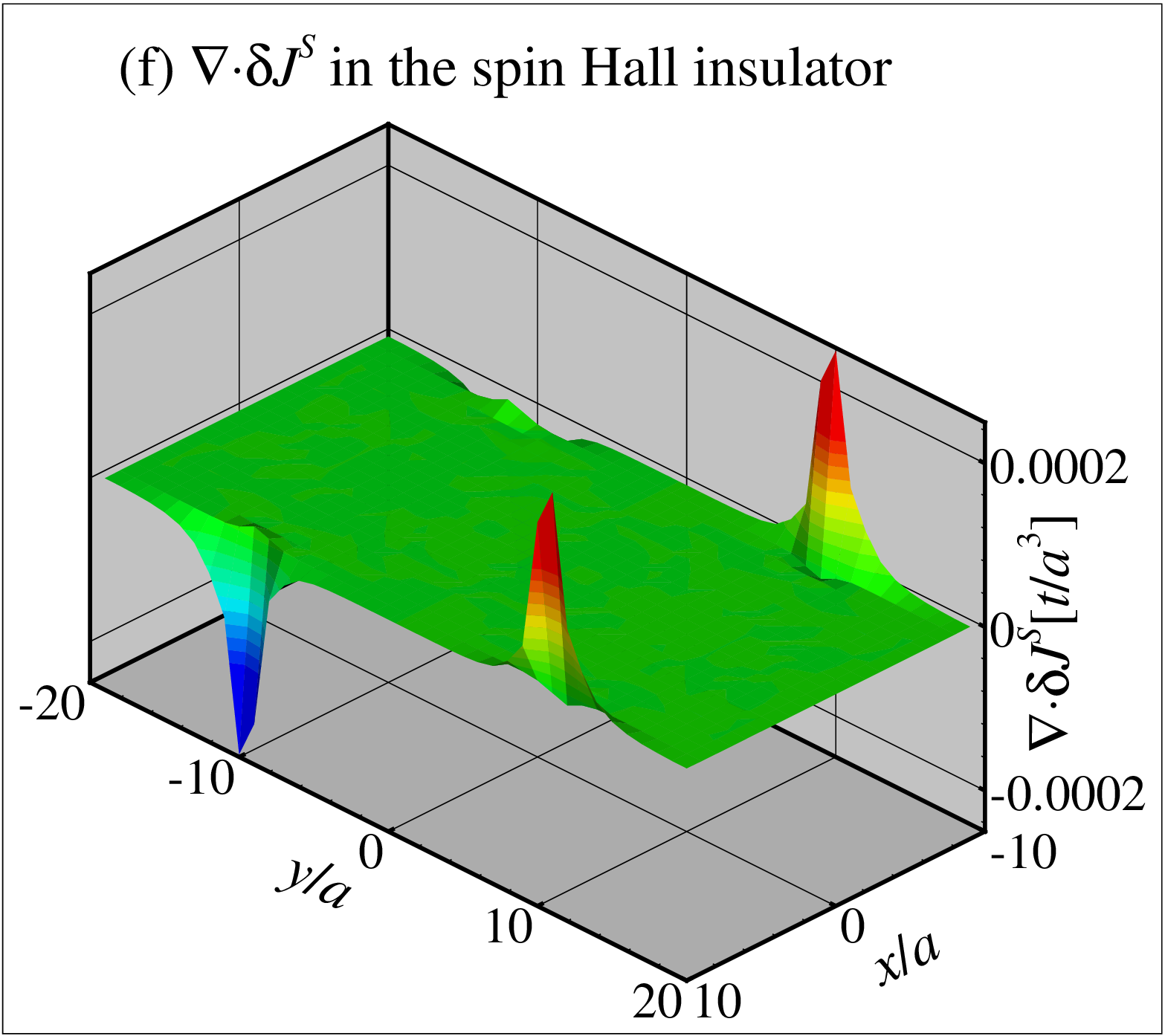}
\caption{
Local density of states per unit cell 
at the Fermi energy $N(E_{F})$,
spin current $\delta\bm{J}^{S_{z}}$ and
the divergence of spin current $\bm{\nabla}\cdot\delta\bm{J}^{S_{z}}$
for the QSHS and the SHI with $L_{x}\times L_{y}=20a\times 40a$.
The electrodes are attached at $\pm L_{x}/2$
from $-L_{y}/4$ to $L_{y}/4$.
The chemical potential of electrons is $\mu_{0}+\delta\mu/2$
at $x=-L_{x}/2$ and $\mu_{0}-\delta\mu/2$ at $x=L_{x}/2$.
As for the QSHS,
the charge current $\delta\bm{J}$ (not shown) flows along the edges.
}
\label{fig:LDOS-dJs}
\end{figure}

The self-consistent equations are solved numerically 
for the system of finite size $L_{x}\times L_{y}$. 
The setup is shown in Fig.~\ref{fig:LDOS-dJs}, where the electrodes are 
attached at $x = \pm L_{x}/2$,
and we take the disorder strength as $\gamma = 1$.
For the moment, we impose the open boundary condition in the $y$-direction.
The difference of chemical potential at the two electrodes are fixed at 
$\delta\mu/L_{x} = 5\times10^{-4}t/a$, which is in the linear
response regime. Here we take the definition of the most naive
spin current as 
$
\bm{J}^{S_{\mu}}_{\bm{r}\bm{r}'} = \frac{1}{2}
\left(S_{\mu}\bm{J}_{\bm{r}\bm{r}'}+\bm{J}_{\bm{r}\bm{r}'}S_{\mu}\right)
$,
where $\bm{S}$ is the spin-$\frac{3}{2}$ matrices, 
and $\bm{J}_{\bm{r}\bm{r}'}$ is the charge current.
This spin current is not conserved in systems with SOI.
Therefore there appears 
the source and sink of the spin current represented by 
$\bm{\nabla}\cdot\delta \bm{J}^{S_{\mu}}$.
 Figure~\ref{fig:LDOS-dJs} shows the bird-view
of the local density of states at the Fermi energy $N(E_{F})$,
the spin current $\delta \bm{J}^{S_{\mu}}$ and 
its divergence $\bm{\nabla}\cdot\delta \bm{J}^{S_{\mu}}$
for (a),(c),(e) the QSHS ($M=-0.5t$, $\Delta=0$, $\mu_{0} = 0$)
and (b),(d),(f)the SHI ($M= 0$, $\Delta=-0.5t$, $\mu_{0} = 0$), respectively,
where $\mu_{0}$ is the chemical potential in equilibrium.
It is clearly seen that the spin current is flowing near the 
electrodes at $x=\pm L_{x}/2$ in both cases.
Here there is a crucial difference between QSHS and SHI.
For QSHS, the density of states $N(E_{F})$ 
due to the massless edge modes is finite around
the whole edges of the sample (Fig.~\ref{fig:LDOS-dJs}(a)),
while it is produced only through the hybridization
between the states of the electrode and
the conduction/valence bands of SHI (Fig.~\ref{fig:LDOS-dJs}(b)).
Correspondingly, the spin current is terminated at the two ends of
each electrode ($x=\pm L_{x}/2$, $y=\pm L_{y}/4$)
with the source and sink of the spin current in SHI,
because the edge modes are gapped (Fig.~\ref{fig:LDOS-dJs}(d),(f)).
On the other hand, the spin current in QSHS 
is flowing from the edges at $x=\pm L_{x}/2$
to the other edges $y= \pm L_{y}/2$,
although it is not conserved also in this case
(Fig.~\ref{fig:LDOS-dJs}(c),(e)). 

\begin{figure}[hbt]
\includegraphics[scale=0.3]{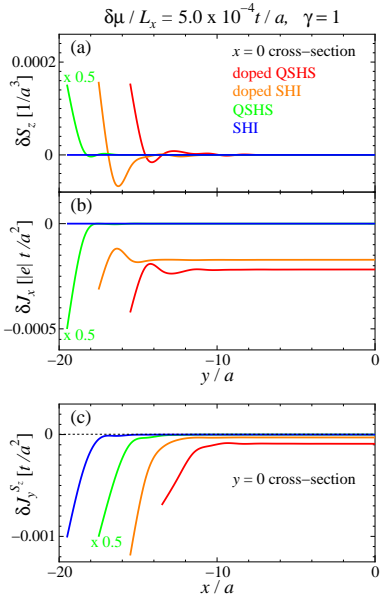}
\caption{
Distribution of (a) $\delta S_{z}$ and (b) $\delta J_{x}$ at $x = 0$
in the system with $L_{x}\times L_{y} = 20a\times 40a$,
and (c) $\delta J^{S_{z}}_{y}$ at $y = 0$
in the system with $L_{x}\times L_{y} = 40a\times 20a$.
The electrodes are attached at $\pm L_{x}/2$
from $-L_{y}/2$ to $L_{y}/2$.
For clarity, only the region of $y < 0$ is shown in each panel,
and $\delta S_z$ is an odd function of $y$.
The curves are shifted in the $y$-direction for 
$\delta S_{z}$ and $\delta J_{x}$
and in the $x$-direction for $\delta J^{S_{z}}_{x}$.
} 
\label{fig:profile}
\end{figure}

Figure~\ref{fig:profile}
shows the details of the deviations from the equilibrium state, i.e.,
(a) the spin $z$-component $\delta S_{z}$ and (b) the charge current 
$x$-component $\delta J_{x}$ in the $x = 0$ cross-section, and 
(c) the spin current $y$-component $\delta J^{S_z}_{y}$
in the $y = 0$ cross-section.
It is noted that the electrodes are attached 
all through the edges at $x=\pm L_{x}/2$, i.e., $|y|<L_{y}/2$.
We take the disorder strength as $\gamma = 1$.
Then, the inverse lifetime is $1/\tau \sim 0.5t$ for the bulk and 
$0.65t$ for the edges of the doped systems ($\mu_{0}=t$),
and $0.4t$ for the edges of the QSHS.
Fixing $\delta\mu/L_{x} = 5\times10^{-4}t/a$,
we have also investigated systems with other sizes and shapes 
which are not shown in Fig.~\ref{fig:profile}.
At least in the regime $\gamma \ge 0.5$, $\delta S_{z}$ and $\delta J_{x}$
has almost no system size- nor shape-dependence.
The $x$-dependences of $\delta S_{z}$ and $\delta J_{x}$ are also negligible.
$\delta J^{S_z}_{y}$ has almost no system size- nor shape-dependence in the bulk region
but scales with $\delta\mu$ near the electrodes.
This property of $\delta J^{S_z}_{y}$ means that
the sample size dependence of the spin Hall conductance is little.
The spin accumulation occurs in every doped case and the QSHS.
This is because the dissipation which breaks the time-reversal symmetry
is caused by the charge current and the associated particle-hole 
excitations in bulk in the doped case,
and at the massless edge modes in the QSHS. It is forbidden in SHI, and
hence there appears no spin accumulation, although the 
spin current flows near the contacts with electrodes as described above.

\begin{figure}[hbt]
\includegraphics[scale=0.25]{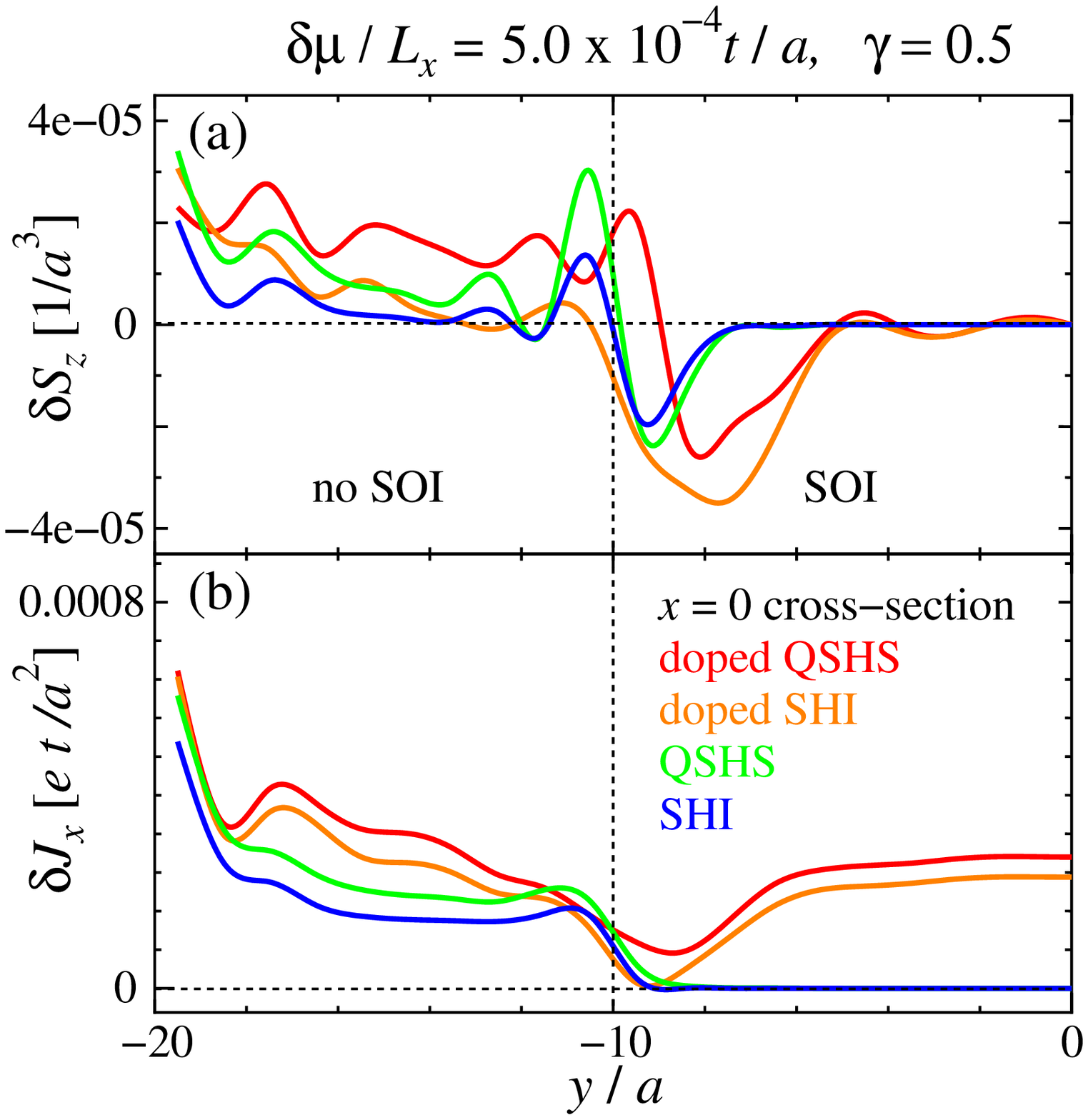}
\includegraphics[scale=0.25]{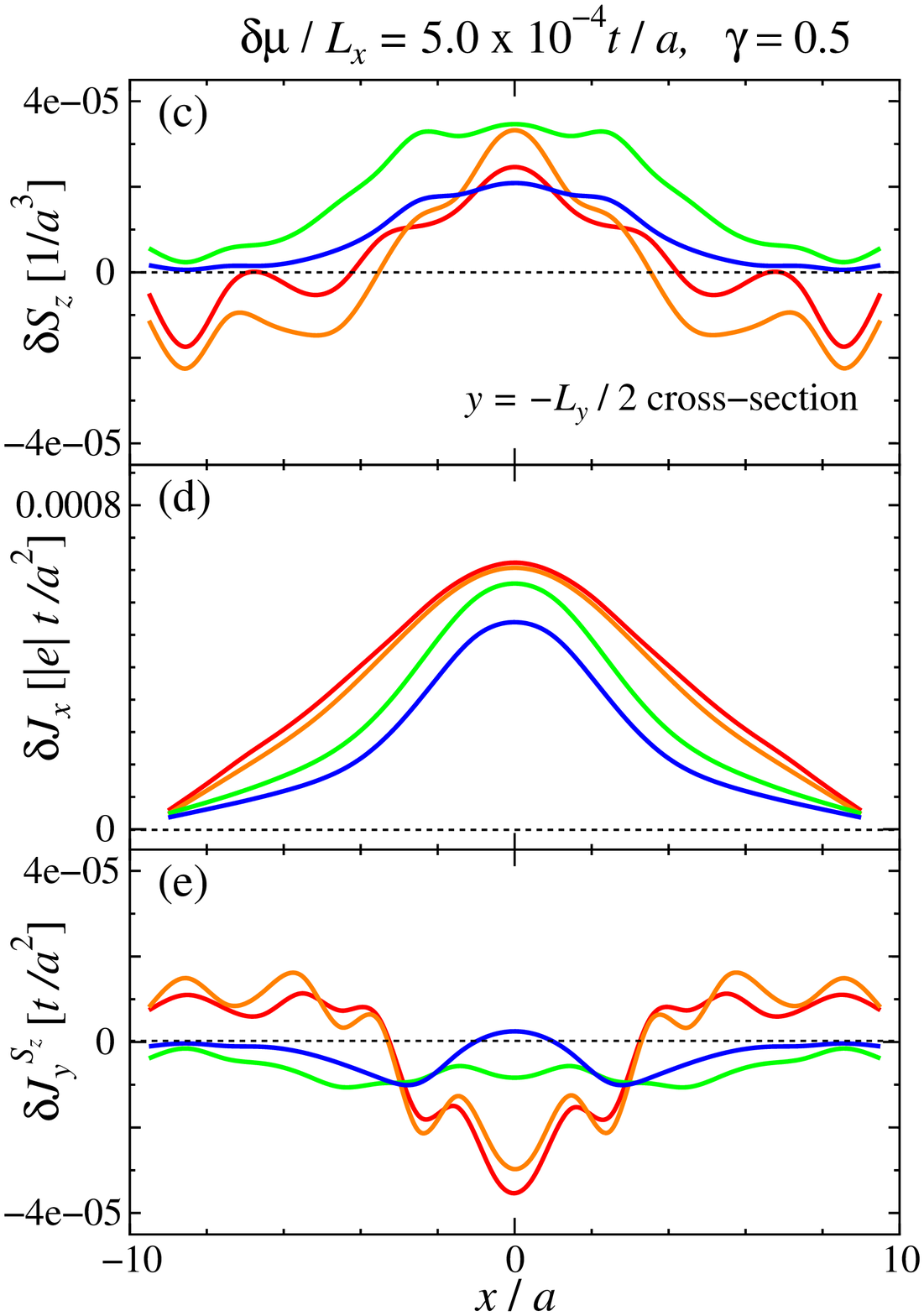}
\caption{
Distribution of (a) $\delta S_z$ and (b) $\delta J_{x}$ at $x = 0$,
(c) $\delta S_z$ and (d) $\delta J_{x}$ at $y=-(L_{y}+L_{y}^{\mathrm{cond}})/2$,and (e) $\delta J^{S_z}_{y}$ 
at a little inside from $y=-(L_{y}+L_{y}^{\mathrm{coond}})/2$
in heterostructures.
For clarity, only the region of $y< 0$ is shown in (a) and (b),
and $\delta S_z$ is an odd function of $y$.
See also Fig.~\ref{fig:QHS-QSHS-SHI}.
} 
\label{fig:profile.hetero}
\end{figure}

Now we consider the case where conductors without SOI are attached
at the edges of the sample.
Its purpose is to investigate whether the SHI can inject spins 
into conductors without SOI and make them accumulate there.
The middle part ($|y| < L_{y}/2$) of the heterostructure 
is described by the Hamiltonian eq.(1), and the side parts 
($L_{y}/2 < |y| < (L_{y}+L_{y}^{\mathrm{cond}})/2$) are the conductors
represented by the Hamiltonian
$
H 
= \sum_{\bm{r},\bm{r}'} 
c^{\dagger}_{\bm{r}}t^{0}_{\bm{r}\bm{r}'}\Gamma_{0}c_{\bm{r}'}
$,
where $\Gamma_{0}$ is the $4\times4$ unit matrix
and 
$t^{0}_{\bm{r}\bm{r}'} = t$ for $\bm{r} = \bm{r}' \pm a \bm{e}_{x,y}$. 
Here we take the total system size as $L_{x}\times(L_{y}+L_{y}^{\mathrm{cond}})=20a\times(20a+20a)$.
The electrodes are attached only to the sample part, $|y|<L_{y}/2$.
Fig.~\ref{fig:profile.hetero} shows (a) $\delta S_z$ 
and (b) $\delta J_{x}$ along the $x = 0$ cross-section,
(c) $\delta S_z$ in the $y=-(L_{y}+L_{y}^{\mathrm{cond}})/2$ cross-section
and (d) $\delta J^{S_{z}}_{y}$ along the cross-section
a little inside from $y=-(L_{y}+L_{y}^{\mathrm{cond}})/2$.
The inverse lifetime in the side conductors is 
$1/\tau \sim 0.3 t$ for the doped systems ($\mu_{0}=t$)
and $1/\tau \sim 0.44 t$ for the QSHS and the SHI,
for the disorder strength $\gamma = 0.5$ in all cases.
We can clearly see the leakage of spins into the side conductors in all cases. 
The discontinuity at the boundary between sample and side conductors is prominent. 
This is attributed to the difference of the energy dispersion and density 
of states between middle and side parts.
It is noted that spins flow out even in the heterostructure of the SHI.
The amount of the accumulated spins in the side conductors 
 is mainly determined by the leakage in charge current.
The obtained spin accumulation is nearly the same as in the doped case.
Therefore, the SHI is more efficient than the doped system,
because there is no current in the sample in the case of the SHI
and there is no energy loss in that part.
Even in our small system, the energy cost of undoped-SHI
is about 45 $\%$ of that of doped-SHI.
\begin{figure}[hbt]
\includegraphics[scale=0.3]{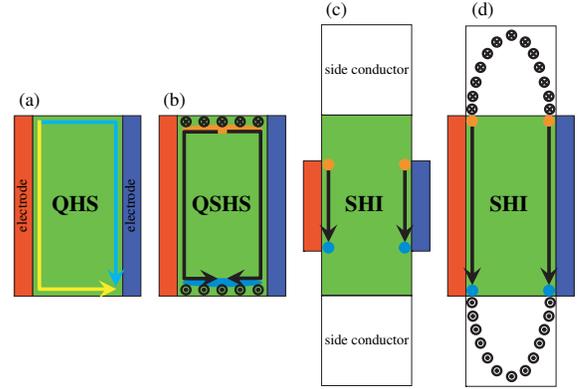}
\caption{
Schematic view of the charge current in (a) QHS,
and the spin current in (b) QSHS,
(c) SHI, and (d) heterostructure of SHI, respectively.
Red and blue boxes represent the electrodes with
$\mu_{0}+\delta\mu/2$ and $\mu_{0}-\delta\mu/2$ respectively.
In the QHS, the yellow and light-blue arrows are the deviation 
from equilibrium of electron flow.
In the QSHS and the SHI, the black arrows are the spin current
and the orange and blue ellipses are source and sink of the spin current.
The circles with dot and cross represent the accumulated up- and down-spins.
} 
\label{fig:QHS-QSHS-SHI}
\end{figure}

In conclusion, we have numerically investigated the spin current and
accumulation due to the intrinsic spin Hall effects
in the spin Hall insulator (SHI) by using Keldysh formalism.
The spin current is flowing near the contacts with electrodes, 
which leads to the spin accumulation in the attached conductors 
associated with the dissipative leakage current
as shown in Fig.~\ref{fig:QHS-QSHS-SHI}.

The authors thank S. Murakami and  B.~K.~Nikoli\'{c} 
for fruitful discussions. 
This work is financially supported by NAREGI Grant, 
Grant-in-Aids from the Ministry of Education,
Culture, Sports, Science and Technology of Japan.


\begin{thebibliography}{99}
\bibitem{MNZ}
S.~Murakami, N.~Nagaosa, and S.-C.~Zhang,
Science {\bf 301}, 1348 (2003).
\bibitem{MNZ1}
S.~Murakami, N.~Nagaosa, and S.-C.~Zhang,
Phys. Rev. B {\bf 69}, 235206 (2004).
\bibitem{Sinova}
J.~Sinova
{\it et al.},
Phys. Rev. Lett. {\bf 92}, 126603 (2004).

\bibitem{SHI}
S.~Murakami, N.~Nagaosa, and S.-C.~Zhang,
Phys. Rev. Lett. {\bf 93}, 156804 (2004) 

\bibitem{Kane}
C.~L.~Kane and E.~J.~Mele, cond-mat/0411737.

\bibitem{Culcer}
D.~Culcer
{\it et al.},
Phys. Rev. Lett. {\bf 93}, 046602 (2004).
\bibitem{PZhang}
P.~Zhang {\it et al.},
cond-mat/0503505.

\bibitem{Wen}
X.~G.~Wen, Phys. Rev. B {\bf 41}, 12838 (1990).

\bibitem{Onoda}
M.~Onoda and N.~Nagaosa,
J. Phys. Spc. Jpn. {\bf 71}, 19 (2002).

\bibitem{Datta}
M.~J.~McLennan, Y.~Lee, and S.~Datta,
Phys. Rev. B {\bf 43}, 13846 (1991).

\bibitem{Nikolic}
B.~K.~Nikoli\'{c}
{\it et al.},
cond-mat/0412595.
\end{thebibliography}
\end{document}